\documentclass[12pt,a4,epsf]{article}
\usepackage{graphicx}
\usepackage{amssymb}
\global\arraycolsep=2pt
\input{epsf}
\usepackage{graphicx}
\def\@fmsl@sh#1#2#3{\m@th\ooalign{$\hfil#1\mkern#2/\hfil$\crcr$#1#3$}}
 \def\eq#1\en{\begin{equation}#1\end{equation}}
\def\s[#1,#2]{[#1\stackrel{\star}{,}#2]}
\def\sx[#1,#2]{[#1\stackrel{\star_{x}}{,}#2]} 
\begin{document}
\makeatletter
\def\fmslash{\@ifnextchar[{\fmsl@sh}{\fmsl@sh[0mu]}}
\def\fmsl@sh[#1]#2{%
   \mathchoice
     {\@fmsl@sh\displaystyle{#1}{#2}}%
     {\@fmsl@sh\textstyle{#1}{#2}}%
     {\@fmsl@sh\scriptstyle{#1}{#2}}%
     {\@fmsl@sh\scriptscriptstyle{#1}{#2}}}
\def\@fmsl@sh#1#2#3{\m@th\ooalign{$\hfil#1\mkern#2/\hfil$\crcr$#1#3$}}
\makeatother

\thispagestyle{empty}
\begin{titlepage}
\begin{flushright}
\end{flushright}

\vspace{0.3cm}
\boldmath
\begin{center}
   \Large {\bf Dynamics of the Fisher Information Metric}
\end{center}
\unboldmath
\vspace{0.8cm}
\begin{center}
   {\large Xavier
Calmet\footnote{email:calmet@physics.unc.edu}$^\dagger$ and Jacques
Calmet\footnote{
email:calmet@ira.uka.de}$^*$}\\ \end{center}
\begin{center}
  {\sl
$^\dagger$University of North Carolina at Chapel Hill, Chapel Hill,
NC 27599, USA}
\\
{\sl $^*$Institute for Algorithms and Cognitive Systems (IAKS)\\
University of Karlsruhe (TH), D-76131 Karlsruhe, Germany}
\end{center}
\vspace{\fill}

\begin{abstract}
\noindent
We present a method to generate probability
distributions that correspond to metrics obeying partial differential
equations generated by extremizing a functional
$J[g^{\mu\nu}(\theta^i)]$, where $g^{\mu\nu}(\theta^i)$ is the Fisher
metric. We postulate that this functional of the dynamical variable
$g^{\mu\nu}(\theta^i)$  is stationary with respect to small
variations of these variables. Our approach enables a dynamical
approach to Fisher information metric. It allows to impose symmetries
on a statistical system in a systematic way. This work is mainly 
motivated by the entropy approach to nonmonotonic reasoning.
\end{abstract}
\end{titlepage}
\section{Introduction}

Nonmonotonic reasoning is basically a static process. The question
whether it is possible to introduce some sort of dynamics in a
reasoning process is open. A motivation would be that nonmonotonic
reasoning leads to solutions of classification and degree of belief
problems.  Dynamics would allow to consider or introduce symmetries
into these games in a very systematic way. A question is thus to
assess whether we can define dynamical reasoning processes and how.
The seminal paper \cite{Pearl} presented a maximum entropy approach to
nonmonotonic reasoning. In \cite{Stachniak} Stachniak points out the
link between universal algebras and AI with a special emphasis to
nonmonotonic reasoning. There is a close connection between
satisfiability and preferential matrix representation ideas.  The
matrix representation arises from a set of equations that are
weighted. An open problem is to give a meaning to these weights.
Furthermore, this approach is in fact closely linked to the definition
of the problem in terms of many valued logics when reasoning is seen
as an inference process.  This very simple survey leads to a first
track i.e. entropy. Indeed Fisher information metric \cite{Fisher} is
an expression for entropy and it is expressed as a matricial equation
that can be linked to the semantic defined by Stachniak. Then, we can
use tools from Physics to introduce dynamical features into the Fisher
metric formalism.  Although these tools are well-known from any
physicists, it looks like such a study was never completed.  We have
mentioned only nonmonotonic reasoning since it is closer to the
research interests of one of the authors, but we could have quoted
quantum computing where dynamical features of entropy ought to bring
new ideas.  In this paper we concentrate on the theoretical problem to
introduce dynamics into the Fisher formalism.

The Fisher information metric can be calculated once a probability
distribution has been chosen. In this work we wish to present a method
to generate probability distributions that correspond to metrics
obeying partial differential equations (Euler-Lagrange PDEs) generated
by extremizing a functional $J[g^{\mu\nu}(\theta^i)]$, where
$g^{\mu\nu}(\theta^i)$ is the Fisher metric. We will postulate that
this functional of the dynamical variable $g^{\mu\nu}(\theta^i)$ is
stationary with respect to small variations of these variables. Our
approach enables a dynamical approach to Fisher information metric. It
allows to impose symmetries on a statistical system in a systematic
way. We will show how to obtain partial differential equations for the
probability distributions. Imposing that the functional remains
invariant under certain transformations of the coordinates $\theta^i$
will constrain the class of probability distributions.  These
symmetries can also be broken in a consistent way using methods
borrowed from theoretical particle physics and general relativity. As an
example we will require that the functional is covariant under general
transformations of the coordinates $\theta$. This will lead to
non-linear partial differential equations corresponding to Einstein's
equations \cite{Einstein:1916vd} of general relativity (see e.g.
\cite{Weinberg} for a nice introduction to general relativity). The
requirement of that the functional remains invariant under
infinitesimal general coordinate transformations $\theta^i \to
{\theta^i}'$ leads to Bianchi identities.

A set of distributions $p_{\theta}=p(\theta)$ parametrized by
$\theta^i$ with $i \in \{1...n\}$ is a manifold. The Riemanian metric
on this manifold is Fisher information metric defined by:
\begin{eqnarray}
g_{\mu\nu}=\int_X d^4\!x p_\theta(x) \left
(\frac{1}{p_\theta(x)} \frac{\partial p_\theta(x)}{\partial
\theta^\mu} \right )
  \left
(\frac{1}{p_\theta(x)}
\frac{\partial p_\theta(x)}{\partial \theta^\nu} \right ).
\end{eqnarray}
In the sequel we shall first briefly introduce the concept of Fisher
information metric insisting on the connection to the concept of
Shannon entropy and of the Kullback-Leibler distance. We shall then
introduce the dynamics in the Fisher information metric. We shall
argue that requiring the covariance of the functional
$J[g^{\mu\nu}(\theta^i)]$ under general transformations of $\theta^i$
is natural since the Fisher information metric is itself invariant
under reparametrization of the manifold \cite{Corcuera}. We note that
the Fisher information metric is also invariant under transformations
of the random variable $x \in X$. We will concentrate on general
coordinate invariance, but in principle depending on the application,
any type of symmetry could be imposed on the functional.

\section{Review of Fisher information metric}

There are many excellent reviews and books on Fisher information
metric, a nice introduction can be found in \cite{Frieden}. Fisher
information metric has been applied in different fields. This concept
appears in such different fields as
e.g.  instanton calculus \cite{Yahikozawa:2003ij}, ontology
\cite{jcalmet}, models for short distance modifications of space-time
\cite{jcalmet2} or econometrics \cite{Marriott}. Symbolic computations
of Fisher information matrices have also been considered
\cite{Peeters}. We shall give a brief derivation of the information
metric.
A distance $d(P_1,P_2)$ between two points $P_1$ and $P_2$ has to
satisfy the
following three axioms:
\begin{itemize}
\item[1.] Positive definiteness: $\forall P_1, P_2: d(P_1,P_2) \ge 0$
\item[2.] Symmetry $d(P_1,P_2)=d(P_2,P_1)$
\item[3.] Triangle inequality: $\forall P_1, P_2, P_3: d(P_1,P_2) \le
d(P_1,P_2) +d(P_1,P_3)$.
\end{itemize}

It is often useful to introduce a concept of distance between elements
of a more abstract set.  For example, one could ask what is the
distance between two distributions between e.g. the Gaussian and
binomial distributions. It is useful to introduce the concept of
entropy as a mean to define distances. In information theory,
Shannon entropy \cite{Shannon} represents the information content of
a message or, from the receiver point of view, the uncertainty about
the message the sender produced prior to its reception. It is defined
as
\begin{eqnarray}
- \sum_i p(i) \log p(i),
\end{eqnarray}
where $p(i)$ is the probability of receiving the message $i$. The unit
used is the bit. The relative entropy can be used to define a
``distance'' between two distributions $p(i)$ and $g(i)$. The
Kullback-Leibler \cite{Kullback} distance or relative entropy is
defined as
\begin{eqnarray}
D(g||p)&=& \sum_i g(i) \log \frac{g(i)}{p(i)}
\end{eqnarray}
where $p(i)$ is the real distribution and $g(i)$ is an assumed
distribution. Clearly the Kullback-Leibler relative entropy is not a
distance in the usual sense: it satisfies the positive definiteness
axiom, but not the symmetry or the triangle inequality axioms. It is
nevertheless useful to think of the relative entropy as a distance
between distributions.

The Kullback-Leibler distance is relevant to discrete sets. It can be
generalized to the case of continuous sets. For our purposes, a
probability distribution over some field (or set) $X$ is a
distribution $p:X \in \mathbb{R}$, such that
\begin{itemize}
\item[1.] $\int_X d^4\!x \  p(x)=1$
\item[2.] For any finite subset $S\subset X$, $\int_S d^4\!x \
p(x)>0$.
\end{itemize}
We shall consider families of distributions, and parametrize them by
a
set of continuous parameters $\theta^i$ that take values in some open
interval $M \subseteq \mathbb{R}^4$. We use the notation $p_\theta$ to
denote members of the family. For any fixed $\theta$, $p_\theta: x
\mapsto p_\theta(x)$ is a mapping from $X$ to $\mathbb{R}$. We shall
consider the extension of the family of distributions $F=\{ p_\theta|
\theta \in M\}$, to a manifold ${\cal M}$ such that the points $p\in
{\cal M}$ are in one to one correspondence with the distributions
$p\in F$. The parameters $\theta$ of $F$ can thus be used as
coordinates on ${\cal M}$.

The Kullback number is the generalization of the Kullback-Leibler
distance for continuous sets. It is defined as
\begin{eqnarray}
I(g_\theta|| p_\theta)&=&
\int d^4\!x  g_\theta(x) \log \frac{g_\theta(x)}{p_\theta(x)}.
\end{eqnarray}
Let us now study the case of an infinitesimal difference between
$q_\theta(x)=p_{\theta +\epsilon v}(x)$ and $p_\theta(x)$:
\begin{eqnarray}
I(p_{\theta +\epsilon v}||p_\theta)&=&  \int d^4\!x  p_{\theta
+\epsilon v}(x)
\log \frac{p_{\theta +\epsilon v}(x)}{p_\theta(x)}.
\end{eqnarray}
Expanding in $\epsilon$ and keeping $\theta$ and $v$ fix one finds:
\begin{eqnarray}
I(p_{\theta +\epsilon v}||p_\theta)&=&I(p+\epsilon||p)
\vert_{\epsilon=0}+ \epsilon \ I^\prime(\epsilon)\vert_{\epsilon=0}+
\frac{1}{2}\epsilon^2 \
I^{\prime\prime}(\epsilon)\vert_{\epsilon=0}+{\cal O}(\epsilon^3).
\end{eqnarray}
One finds $I(0)=I^\prime(0)=0$ and
\begin{eqnarray}
I^{\prime\prime}(0)&=& v^\mu \left ( \int_X d^4\!x p_\theta(x) \left
(\frac{1}{p_\theta(x)} \frac{\partial p_\theta(x)}{\partial
\theta^\mu} \right )
  \left
(\frac{1}{p_\theta(x)}
\frac{\partial p_\theta(x)}{\partial \theta^\nu} \right ) \right )
v^\nu.
\end{eqnarray}
We can now identify the Fisher information metric \cite{Fisher} on a
manifold of probability distributions as
\begin{eqnarray} \label{Fishermetric}
g_{\mu\nu}=\int_X d^4\!x p_\theta(x) \left
(\frac{1}{p_\theta(x)} \frac{\partial p_\theta(x)}{\partial
\theta^\mu} \right )
  \left
(\frac{1}{p_\theta(x)}
\frac{\partial p_\theta(x)}{\partial \theta^\nu} \right ).
\end{eqnarray}
It has been show that this matrix is a metric on a manifold of
probability distributions, see e.g.  \cite{rodriguez1}. Corcuera and
Giummol\`e \cite{Corcuera} have shown that the Fisher information
metric is invariant under reparametrization of the sample space $X$
and that it is covariant under reparametrizations of the manifold,
i.e. the parameter space, see e.g. \cite{Wagenaar} for a
review.

\section{Dynamics of Fisher information metric}
Given a distribution $p(\theta)$, the Fisher information metric can
be calculated using (\ref{Fishermetric}). Here we wish to approach this
issue from a different point of view and derive  the Fisher
information metric from a dynamical perspective. We introduce a
functional $J[g^{\mu\nu}(\theta^i)]$. To constrain the functional
dependence on $g^{\mu\nu}(\theta^i)$, we impose that this functional
be invariant under general transformations of the coordinates
$\theta^i$. As already mentioned, the Fisher information
metric is covariant under reparametrizations of the manifold, i.e. it
is covariant under general coordinate transformations  $\theta^i \to
{\theta^i}'$. It thus seems natural to posit that the functional
$J[g_{\mu\nu}]$, describing the dynamics of the metric, is invariant
under general coordinate transformations  $\theta^i \to {\theta^i}'$.
We shall first introduce a few basic quantities which are well known
from differential geometry and the general theory of relativity.

\subsection{Definitions}
Differential geometry is the tool we shall be using, just as in
general relativity, the concepts of differential geometry we need are
nicely introduced in \cite{Weinberg}. Let us define a few quantities.
The affine connection is defined as:
\begin{eqnarray}
\Gamma^\sigma_{\lambda \nu}= \frac{1}{2} g^{\nu \sigma}
  \left (
\frac{\partial g_{\mu \nu}}{\partial \theta^\lambda}
+
\frac{\partial g_{\lambda \nu}}{\partial \theta^\mu}
-
\frac{\partial g_{\mu \lambda}}{\partial \theta^\nu}
\right ).
\end{eqnarray}
The curvature tensor is given by:
\begin{eqnarray}
R^\lambda_{\mu\nu\kappa}&=& \frac{\partial
\Gamma^\lambda_{\mu\nu}}{\partial \theta^\kappa}
-  \frac{\partial \Gamma^\lambda_{\mu\kappa}}{\partial \theta^\nu}
+\Gamma^\eta_{\mu\nu} \Gamma^\lambda_{\kappa\eta}
-\Gamma^\eta_{\mu\kappa} \Gamma^\lambda_{\nu \eta}
\end{eqnarray}
it is the only tensor that can be constructed from the metric tensor
$g_{\mu \nu}$ and its first and second derivatives. We shall also
need the Ricci tensor which reads
\begin{eqnarray} \label{ricci}
R_{\mu\kappa} &=&R^{\lambda}_{\mu\lambda\kappa}
\end{eqnarray}
and the curvature scalar is given by
\begin{eqnarray}\label{curv}
R &=&g^{\mu\kappa}R_{\mu\kappa}.
\end{eqnarray}

\subsection{Dynamics}
We now have all the necessary tools to introduce the invariant
functional which describes the dynamics of Fisher information metric.
Because of \cite{Corcuera}, it seems
  natural to posit  that the functional $J[g_{\mu\nu}]$, describing the dynamics of
the metric, is invariant under general coordinate transformations
$\theta^i \to {\theta^i}'$\footnote{this situation arises in
Eintein's theory of  general relativity.}.
We shall start from the functional:
\begin{eqnarray}
J[g_{\mu\nu}]= \frac{-1}{16 \pi} \int \sqrt{g(\theta)} R(\theta)
d^4\theta
\end{eqnarray}
where $g=$det$g^{\mu\nu}$ and $R(\theta)$ is the curvature scalar.
This functional is a scalar and it is invariant under general
coordinate transformations. Note that $R(\theta)$ is a scalar and is
thus invariant under general coordinate transformations, $d^4\theta$
transforms as
\begin{eqnarray}
d^4\theta' &\to& \left | \frac{\partial \theta'}{\partial \theta}
\right | d^4 \theta
\end{eqnarray}
and $\sqrt{g(\theta)}$ transforms as:
\begin{eqnarray}
\sqrt{g(\theta')} &\to & \left | \frac{\partial \theta}{\partial
\theta'} \right | \sqrt{g(\theta)}.
\end{eqnarray}
The functional is a scalar and  thus invariant under general
coordinate transformations.

The variation of  $J[g_{\mu\nu}]$  with respect to $g_{\mu\nu}$ leads
to
\begin{eqnarray}
\delta J[g_{\mu\nu}] = \frac{1}{16 \pi} \int \sqrt{g} \left[
R^{\mu\nu}(\theta)-\frac{1}{2} g^{\mu\nu} R(\theta) \right ] \delta
g_{\mu\nu} d^4 \theta,
\end{eqnarray}
where $R^{\mu\nu}(\theta)$ is the Ricci tensor introduced in
eq.(\ref{ricci}) and $R(\theta)$ is the curvature scalar (see eq.
(\ref{curv})).
The requirement that $J[g_{\mu\nu}]$ is invariant with respect to
variation of the metric $g_{\mu\nu}$, $\delta g_{\mu\nu}$ leads to
the following partial differential equations (the Euler-Lagrange
equations):
\begin{eqnarray} \label{Einsteinvide}
R^{\mu\nu}(\theta)-\frac{1}{2} g^{\mu\nu}(\theta)  R(\theta)=0.
\end{eqnarray}
Contracting eq. (\ref{Einsteinvide}) with $g_{\mu\nu}$,  one finds
$R=0$, the  partial differential equations (the Euler-Lagrange
equations) are thus:
  \begin{eqnarray} \label{Einstein}
R^{\mu\nu}(\theta)=0
\end{eqnarray}
which correspond to the equations obtained by Einstein in his theory
of general relativity \cite{Einstein:1916vd}. If the statistical
system under consideration is invariant under reparametrization of
the manifold, the dynamics of Fisher information metric is governed
by these  partial differential equations.

Another interesting case is the one of infinitesimal transformations
$\theta=\theta+\epsilon(\theta)$. The functional $J[g_{\mu\nu}]$
being a scalar is also invariant under this class of transformation.
This transformation of $\theta$ implies the following variation for
the metric:
\begin{eqnarray}
\delta g_{\mu\nu}(\theta)&=&-g_{\mu\lambda}(\theta)\frac{\partial
\epsilon^\lambda(\theta)}{\partial \theta^\nu}
-g_{\lambda \nu}(\theta)\frac{\partial
\epsilon^\lambda(\theta)}{\partial \theta^\mu}
-\frac{\partial g_{\mu\nu}(\theta)}{\partial \theta^\lambda}
\epsilon^\lambda(\theta).
\end{eqnarray}
The variation of the functional $J$ with respect to this
transformation leads to the contracted Bianchi identity:
\begin{eqnarray}
[R^\nu_{\ \lambda} -\frac{1}{2} \delta^\nu_{\ \lambda} R]_{;\nu}=0;
\end{eqnarray}
where the semicolon stands for  the covariant derivative which is defined by:
\begin{eqnarray}
V_{\mu;\lambda} = \frac{\partial{V_\mu}}{\partial \theta^\lambda}-
\Gamma^\lambda_{\mu \nu} V^\nu.
\end{eqnarray}

It is possible to introduce more dynamics in the model by introducing
another source (the partial differential equations being nonlinear,
the metric is a source for itself already) for the metric in the form
of a symmetric tensor $T^{\mu\nu}$ which transforms as a
contravariant tensor (i.e. $T^{\mu\nu'}\to\frac{\partial
\theta^\mu}{\partial \theta^\rho} \frac{\partial\theta^\nu}{\partial
\theta^\sigma} T^{\rho \sigma}$). The modified functional then
becomes:
\begin{eqnarray}
J[g_{\mu\nu}]= \frac{-1}{16 \pi} \int \sqrt{g(\theta)} R(\theta)
d^4\theta+\frac{1}{2}
\int \sqrt{g(\theta)} T^{\mu\nu} g_{\mu\nu} d^4\theta.
\end{eqnarray}
This functional leads to the partial differential equations:
\begin{eqnarray}
R^{\mu\nu}(\theta)-\frac{1}{2} g^{\mu\nu}(\theta) R(\theta)+ 8 \pi
T^{\mu\nu}(\theta) =0.
\end{eqnarray}
In the general theory of relativity the case $T^{\mu\nu}(\theta)=0$
correspond to empty space. In a statistical system case, the tensor
$T^{\mu\nu}(\theta)$ can be used to implement constraints on the
statistical system under consideration. Notice that symmetries, in
our case general coordinate invariance, can be broken by generating
terms in $T^{\mu\nu}$ that are not transforming as second rank
tensors. Imposing symmetries on a system and breaking these
symmetries usually leads to relations between the parameters of the
model. This can be very useful for applications.

Obviously the functional we have chosen is the simplest case
possible. We have chosen to build the curvature tensor using only the
metric and its first and second derivative. One could obviously
obtain more complicated models by taking higher derivatives into
account. One could also introduce further dynamics in the model. As
an example one can add a scalar $\phi$ in the theory and consider  a
so-called  Brans-Dicke model \cite{Brans:1961sx}:
\begin{eqnarray}
J[g_{\mu\nu}]= \frac{-1}{16 \pi} \int \sqrt{g(\theta)} \phi(\theta)
R(\theta) d^4\theta + \mbox{dynamics for $\phi(\theta)$}.
\end{eqnarray}
This functional is also a scalar i.e. it is invariant under general
coordinate transformations, but it leads to a different set of
partial differential equations.

\subsection{Constraints for the probability distributions}
If one inserts the expression for the Fisher information metric
(\ref{Fishermetric}) in the partial  differential equations
(\ref{Einstein}) obtained postulating that the functional is
invariant under general coordinate transformations (note that
$R_{\mu\nu}$ and $R$ are functions of $g_{\mu \nu}$), one finds a
lengthy and complicated expression. It is a differential equations
for the distribution $p_\theta(x)$. It is best to first find
solutions to the partial differential equations  (\ref{Einsteinvide})
or  (\ref{Einstein}) to obtain the metric constrained by the
symmetries of the problem  and then to deduce the constraints for the
distribution $p_\theta(x)$.
Finding an exact solution to the partial differential equations
(\ref{Einsteinvide}) or  (\ref{Einstein})  is in general very
difficult, however when the problem under consideration has enough
symmetries, certain solutions are known.

Obviously, a trivial diagonal metric with constant entries of the
type $(1,...,1)$ is a solution of the partial differential equations
(\ref{Einsteinvide}). One obtains the following constraints for
probability distributions:
\begin{eqnarray}
\int_X d^4\!x p_\theta(x) \left
(\frac{1}{p_\theta(x)} \frac{\partial p_\theta(x)}{\partial
\theta^\mu} \right )
  \left
(\frac{1}{p_\theta(x)}
\frac{\partial p_\theta(x)}{\partial \theta^\mu} \right )&=&1, \\
\nonumber
\int_X d^4\!x p_\theta(x) \left
(\frac{1}{p_\theta(x)} \frac{\partial p_\theta(x)}{\partial
\theta^\mu} \right )
  \left
(\frac{1}{p_\theta(x)}
\frac{\partial p_\theta(x)}{\partial \theta^\nu} \right )&=&0 \
{\mbox{ if $\mu \neq \nu$}}.
\end{eqnarray}
this constraint is trivial and for example Gaussian distributions
fulfill it.

Another example is that of a four dimensional problem which is
spherical symmetric in three coordinates  (isotropic) say
$\theta^1,\theta^2$ and $\theta^3$ and with a 
metric that is independent on the fourth coordinate $\theta^0$. In that case a
solution was found by Schwarzschild \cite{Schwarzschild:1916ae}  (see
also e.g. \cite{Weinberg}). Using  spherical coordinates for three
coordinates $\theta^1,\theta^2$ and $\theta^3$,   the metric is of
the form $(1-\frac{2 \alpha}{r}, \left( 1-\frac{2
\alpha}{r}\right)^{-1}, r^2, r^2 \sin^2 \theta)$ for a four
dimensional problem denoting the coordinates by $(\tau,r, \theta,
\phi)$, and where $\alpha$ is an integration constant. We thus obtain
the following constraints for a probability distribution:
\begin{eqnarray}
\int_X d^4x
\frac{1}{p_\theta(x)}  \left (\frac{\partial p_\theta(x)}{\partial
\tau} \right )^2
   &=&1-\frac{2 \alpha}{r} \\
  \int_X d^4x
\frac{1}{p_\theta(x)}  \left (\frac{\partial p_\theta(x)}{\partial r}
\right )^2 
  &=&
\left ( 1-\frac{2 \alpha}{r} \right)^{-1} \nonumber \\
\int_X d^4x
\frac{1}{p_\theta(x)}  \left (  \frac{1}{r}\frac{\partial p_\theta(x)}{\partial
\theta} \right )^2
&=&r^2 \nonumber \\
\int_X d^4x
\frac{1}{p_\theta(x)}  \left ( \frac{1}{r \sin\theta} \frac{\partial p_\theta(x)}{\partial
\phi} \right )^2
&=&r^2 \sin^2 \theta  \nonumber \\
\int_X d^4\!x p_\theta(x) \left
(\frac{1}{p_\theta(x)} \frac{\partial p_\theta(x)}{\partial
\theta^\mu} \right )
  \left
(\frac{1}{p_\theta(x)}
\frac{\partial p_\theta(x)}{\partial \theta^\nu} \right )&=&0 \
{\mbox{ if $\mu \neq \nu$}}.
\end{eqnarray}
This example illustrates how to obtained non-trivial constraints on
the probability distributions.

\section{Conclusions}
In this work we have presented a method to generate probability
distributions that correspond to metrics obeying partial differential
equations generated by extremizing a functional
$J[g^{\mu\nu}(\theta^i)]$, where $g^{\mu\nu}(\theta^i)$ is the Fisher
metric. We have postulated that this functional of the dynamical
variable $g^{\mu\nu}(\theta^i)$  is stationary with respect to small
variations of these variables. Our approach enables a dynamical
approach to Fisher information metric. It allows to impose symmetries
on a statistical system in a systematic way. We have presented
different models and some solutions to these partial differential
equations. There is a very nice analogy between Fisher information
metric and the Einstein's theory of general relativity.  We have
argued  that since the Fisher information
metric is covariant under reparametrizations of the manifold, i.e. it
is covariant under general coordinate transformations  $\theta^i \to
{\theta^i}'$, it is natural to posit that the functional $J[g_{\mu\nu}]$,
describing the dynamics of the metric, is invariant under general
coordinate transformations  $\theta^i \to {\theta^i}'$. This led us
to the functional that determines the dynamics of our models. As
pointed out at the very beginning of the paper we foresee several
applications domains such as reasoning or quantum computing. There is
an additional application that is under investigation and is a
classical one for stochastical methodologies: classification in
insurance classifications \cite{Hipp}. We expect to refine the classification
process through symmetry considerations.

\subsection*{Acknowledgment}
The work of X.C. was supported in part by the US Department of Energy
under Grant No. DE-FG02-97ER-41036.


\end{document}